\PassOptionsToPackage{unicode}{hyperref}
\PassOptionsToPackage{hyphens}{url}
\documentclass[
  10pt,
  a4paper,
]{article}
\usepackage{xcolor}
\usepackage[margin=1in]{geometry}
\usepackage{amsmath,amssymb}
\usepackage{iftex}
\ifPDFTeX
  \usepackage[T1]{fontenc}
  \usepackage[utf8]{inputenc}
  \usepackage{textcomp} 
\else 
  \usepackage{unicode-math} 
  \defaultfontfeatures{Scale=MatchLowercase}
  \defaultfontfeatures[\rmfamily]{Ligatures=TeX,Scale=1}
\fi
\usepackage{lmodern}
\ifPDFTeX\else
\fi
\IfFileExists{upquote.sty}{\usepackage{upquote}}{}
\IfFileExists{microtype.sty}{
  \usepackage[]{microtype}
  \UseMicrotypeSet[protrusion]{basicmath} 
}{}
\usepackage{setspace}
\makeatletter
\@ifundefined{KOMAClassName}{
  \IfFileExists{parskip.sty}{%
    \usepackage{parskip}
  }{
    \setlength{\parindent}{0pt}
    \setlength{\parskip}{6pt plus 2pt minus 1pt}}
}{
  \KOMAoptions{parskip=half}}
\makeatother
\usepackage{longtable,booktabs,array}
\usepackage{calc} 
\usepackage{etoolbox}
\makeatletter
\patchcmd\longtable{\par}{\if@noskipsec\mbox{}\fi\par}{}{}
\makeatother
\IfFileExists{footnotehyper.sty}{\usepackage{footnotehyper}}{\usepackage{footnote}}
\makesavenoteenv{longtable}
\providecommand{\tightlist}{%
  \setlength{\itemsep}{0pt}\setlength{\parskip}{0pt}}
\usepackage{bookmark}
\IfFileExists{xurl.sty}{\usepackage{xurl}}{} 
\hypersetup{
  pdftitle={The AI Adaptation Gap in Higher Education: Students, Faculty, and Administrative Staff},
  pdfauthor={Yuriy S. Braun; Salavat M. Khafizov},
  hidelinks,
  pdfcreator={LaTeX via pandoc}}

\author{}
\date{}

\begin{document}

\setstretch{1.08}
\section{The AI Adaptation Gap in Higher Education: Students, Faculty,
and Administrative
Staff}\label{the-ai-adaptation-gap-in-higher-education-students-faculty-and-administrative-staff}

\subsection{AI Use, Trust, and Institutional Policy at a University
Specializing in Teacher
Education}\label{ai-use-trust-and-institutional-policy-at-a-university-specializing-in-teacher-education}

\textbf{Yuriy S. Braun}\\
Moscow Pedagogical State University, Institute for the Development of
Digital Education, Chair of Digital Education, Moscow, Russia\\
Email: yus.braun@mpgu.su

\textbf{Salavat M. Khafizov}\\
Independent Researcher, Moscow, Russia\\
Email: salavat.khafizov@gmail.com\\
Corresponding author

\subsection{Abstract}\label{abstract}

The purpose of this study was to analyze patterns of artificial
intelligence (AI) use and attitudes toward AI among students, faculty,
and administrative staff at a large university specializing in teacher
education. The analytical sample comprised 1809 students, 250 faculty
members, and 62 administrative staff members (\emph{N} = 2121). Three
role-adapted 75-item questionnaires covered the frequency and contexts
of AI use, perceived usefulness, trust and control, academic integrity
concerns, responsible-use norms, institutional policy clarity, and
perceived improvement in output quality. Data analysis included
descriptive statistics, Welch group comparisons, pooled ordinary least
squares (OLS) models, reliability and dimensionality checks for observed
indices, and exploratory student-only K-means clustering. The results
revealed a pronounced AI adaptation gap across university groups.
Students reported higher current AI-use intensity and perceived
usefulness than faculty and administrative staff, whereas faculty and
administrative staff reported stronger academic integrity concerns and
greater endorsement of responsible-use norms. In the pooled OLS trust
model, perceived usefulness had the strongest standardized positive
association with trust in AI (\(\beta\) = 0.402); institutional policy
clarity also had a positive but weaker association (\(\beta\) = 0.223).
Students reported higher perceived policy clarity than faculty, while
neither group differed significantly from administrative staff.
Exploratory clustering indicated heterogeneity among students in
experience, competence, usefulness, trust, and control, but did not
establish a latent typology across university groups. The
cross-sectional, self-reported data show associations and group
differences rather than causal effects on learning or objective
outcomes.

\textbf{Keywords:} artificial intelligence; higher education; university
students; faculty; educational governance; technology adoption

\subsection{List of abbreviations}\label{list-of-abbreviations}

\begin{itemize}
\tightlist
\item
  AI: artificial intelligence
\item
  ANOVA: analysis of variance
\item
  OLS: ordinary least squares
\item
  PCA: principal component analysis
\end{itemize}

\section{1. Introduction}\label{introduction}

In recent years, generative AI tools have moved from the margins of
technological debate into everyday use at universities. Recent
publications show that the focus has rapidly shifted from broad
discussions of the need for AI in higher education and its potential
affordances to more specific questions of technology adoption, trust,
use profiles, AI literacy, and AI competencies among students, future
teachers, and university faculty (Mo et al., 2026; Lai et al., 2024;
Kasneci et al., 2023; Zawacki-Richter et al., 2019; Tlili et al., 2023;
Cotton et al., 2024; Torres-Díaz et al., 2025; Hackl et al., 2026; Saihi
et al., 2024). This research field already includes studies of student
adoption of AI, differences in AI proficiency between students and
faculty, undisclosed AI use in academic work, and related issues.

However, research on AI in higher education remains uneven: theoretical
discussion is more developed than empirical research on real-world
AI-use contexts among students, faculty, and administrative staff.
Existing empirical studies focus mainly either on students (Stojanov et
al., 2024; Hugerth \& Hugerth, 2026; Lai et al., 2024; Mariñas et al.,
2025) or on faculty (Sun et al., 2026; Burneo-Arteaga et al., 2025),
although broader stakeholder studies are emerging (Humble \& Mozelius,
2026). The groups that jointly shape university practice are rarely
compared directly: students as users in learning contexts, faculty as
actors responsible for pedagogical decisions, and administrative staff
as actors who shape institutional rules, infrastructure, and permissible
modes of use. Yet misalignment among these three stakeholder groups may
shape whether AI is perceived as an educational resource, a convenient
administrative service, or a source of academic and ethical risk.

Research from Russia similarly documents rapid student adoption,
differing student and faculty perceptions, and unresolved questions of
academic integrity and AI-assisted plagiarism (Aleshkovskiy et al.,
2024; Buyakova et al., 2024; Sysoyev, 2024).

This study addresses the gap using data from a large university
specializing in teacher education. In this article, AI is treated not as
a single tool but as a technology embedded in university assignments,
communication, planning, assessment, administrative processes, and
service practices. Following this logic, the analysis examines not only
the frequency of AI use but also how experience, competence, perceived
usefulness, academic integrity concerns, trust, and institutional rules
are associated with current AI-use practices.

The study is guided by the premise that patterns of AI use and trust are
associated not with any single variable but with a combination of four
domains: perceived usefulness, self-reported competence, academic
integrity concerns, and institutional policy clarity. In other words,
university AI-use practices develop not only from the bottom up through
user initiative but also in the context of institutional rules, signals,
and boundaries of permissible use.

Specifically, the study has three interrelated objectives. First, it
describes and directly compares the prevalence of reported AI-service
use, weekly use, and current AI-use intensity among students, faculty,
and administrative staff. Second, it estimates associations among the
observed indices of perceived usefulness, institutional policy clarity,
trust, and current AI-use intensity through regression models. Third, it
uses K-means clustering to describe exploratory heterogeneity in the
student sample. This design supports recommendations differentiated by
stakeholder group and task.

The study addresses the following research questions:

\begin{enumerate}
\def\labelenumi{\arabic{enumi}.}
\tightlist
\item
  To what extent do students, faculty, and administrative staff differ
  in the prevalence of reported AI-service use, weekly use, and current
  AI-use intensity, as well as in perceived usefulness, academic
  integrity concerns, and institutional policy clarity?
\item
  How are perceived usefulness, institutional policy clarity, current
  AI-use intensity, responsible-use norms, and academic integrity
  concerns associated with trust in AI and current AI-use intensity in
  pooled OLS models that include group indicators?
\item
  What exploratory AI-user profiles emerge among students when several
  attitudinal and behavioral measures are considered simultaneously?
\item
  What implications does the identified AI adaptation gap have for
  institutional governance and the development of responsible higher
  education policy?
\end{enumerate}

\section{2. Theoretical Framework}\label{theoretical-framework}

\subsection{2.1. Perceived Usefulness as a Factor in
Integration}\label{perceived-usefulness-as-a-factor-in-integration}

The first analytical domain concerns the perceived usefulness of AI in
higher education. In research on digital technology adoption, perceived
usefulness is usually a central predictor of positive attitudes and
intention to use a technology (Mariñas et al., 2025). This is especially
important in the university context because usefulness operates across
three relevant dimensions: accelerating information search and
processing, improving the quality of feedback, and reducing routine
workload. If users see that AI helps them explain material more
effectively, prepare materials faster, design assignments more clearly,
or distribute administrative work more efficiently, sustained technology
use may become more likely.

In this study, usefulness is understood as a subjective assessment that
AI makes educational or related administrative activity more efficient,
more effective, or more convenient. The same set of AI functions may be
evaluated differently by students, faculty, and administrative staff.
Nevertheless, perceived usefulness is expected to be positively
associated with trust and use intensity, although the cross-sectional
design cannot establish the direction of causal influence.

\subsection{2.2. Risks, Academic Integrity, and Constraints on
Use}\label{risks-academic-integrity-and-constraints-on-use}

The second analytical domain addresses tensions surrounding AI use in
universities: the simplification of academic work coexists with concerns
about a decline in the depth of learning, threats to academic integrity
coexist with productivity gains, and the convenience of automation
coexists with questions of accuracy, privacy, and responsibility (Cotton
et al., 2024; Torres-Díaz et al., 2025). Even when the technology is
perceived as convenient, high perceived risk may be accompanied by a
more cautious attitude toward its use.

For empirical analysis, it is important to distinguish between two
levels of risk. The first is the risk of inaccurate or unreliable
information. The second is the risk of institutionally inappropriate
use, especially in assessment, independent work, and academic integrity.
Only the second level was directly operationalized in the present
analysis; the main models did not include a separate index of risks
related to inaccuracy, privacy, or unreliable information.

\subsection{2.3. Institutional Policy Clarity and
Trust}\label{institutional-policy-clarity-and-trust}

The third domain builds on the idea that, in a rapidly changing
technological environment, users respond not only to the tool itself but
also to the rules that govern its use (Lai et al., 2024). When
university rules are vague or contradictory, uncertainty may increase.
By contrast, a clear policy can specify permitted and restricted uses,
requirements for disclosing AI assistance, and implications for
assessment.

Within this framework, clear rules for AI use should be understood not
merely as a bureaucratic formality but as a potential resource for
institutional governance and pedagogical clarity. For administrative
staff, this is primarily a procedural matter; for faculty, a question of
pedagogical and ethical boundaries; and for students, a question of
understanding permissible conduct. The present study examines this
premise as an observed association between perceived policy clarity and
trust, not as a directional causal effect.

\subsection{2.4. Responsible-Use Norms and Perceived
Quality}\label{responsible-use-norms-and-perceived-quality}

The fourth domain concerns not only how frequently AI is used but also
the quality of that use. AI may become widespread while serving mainly
as a means of completing tasks more quickly but superficially rather
than providing genuine support for learning; it is therefore necessary
to distinguish the use of the technology from norms of responsible use
(Hackl et al., 2026). Responsible use entails checking the accuracy of
information, understanding model limitations, following local rules, and
maintaining reflective control.

In this study, responsible-use norms were operationalized through
disclosure of substantial AI use, explanation of AI-assisted steps, and
judgments about violations in closed-book assessment or undisclosed use.
Responsible-use norms, trust, and perceived improvement in quality were
self-reported measures. They did not measure objective academic
performance and cannot establish that AI use improves or impairs
educational outcomes.

\subsection{2.5. Group Differences}\label{group-differences}

Finally, the study design includes students, faculty, and administrative
staff and assumes that use intensity, perceived usefulness, trust,
risks, and normative attitudes may differ across these groups. The
marked imbalance in group sizes requires methods robust to unequal
variances and sample sizes, and estimates for the administrative group
require particular caution.

\section{3. Method}\label{method}

\subsection{3.1. Study Context and
Design}\label{study-context-and-design}

The study was conducted at a large university specializing in teacher
education and used a cross-sectional survey design. The available
archival metadata identify December 2025 as the data-collection date but
provide no information on the duration of data collection. The analysis
used three cleaned, role-specific datasets corresponding to students,
faculty, and administrative staff. The university's senior
administrative body (the rectorate) served as the institutional approval
body under the university's procedures for research involving human
participants and approved the study protocol and ethics procedures. All
participants were older than 16 years, and informed consent was obtained
before questionnaire completion. The institution remains anonymized in
this manuscript.

The analyzed datasets contained no direct personal identifiers. Columns
for the optional contact item 11.1 were present but contained no values.
Whether any original contact information was stored separately could not
be determined from the available records. The available records did not
document recruitment channels or pilot testing, so those details are not
reported here.

\subsection{3.2. Participants}\label{participants}

The analytical sample comprised 2121 respondents: 1809 students, 250
faculty members, and 62 administrative staff members. Group sizes
correspond to the number of respondents in the final analytical
datasets.

The student questionnaire recorded educational program and year, mode of
study, and age interval. The faculty questionnaire included position,
teaching experience, and age interval, while the administrative
questionnaire included role, length of employment in higher education,
and age interval. Because age was represented by interval categories,
mean ages were not calculated.

Table 1 presents two use measures separately. Reported AI-service use
was defined as reporting use of at least one AI service in item 0.6.
Weekly use was defined as reporting AI use one to three times per week
or more often on role-specific item 1.4 among respondents with a valid
answer to that item.

\textbf{Table 1. Sample composition and two measures of AI use}

\begin{longtable}[]{@{}
  >{\raggedright\arraybackslash}p{(\linewidth - 8\tabcolsep) * \real{0.1579}}
  >{\raggedleft\arraybackslash}p{(\linewidth - 8\tabcolsep) * \real{0.2105}}
  >{\raggedleft\arraybackslash}p{(\linewidth - 8\tabcolsep) * \real{0.2105}}
  >{\raggedleft\arraybackslash}p{(\linewidth - 8\tabcolsep) * \real{0.2105}}
  >{\raggedleft\arraybackslash}p{(\linewidth - 8\tabcolsep) * \real{0.2105}}@{}}
\toprule\noalign{}
\begin{minipage}[b]{\linewidth}\raggedright
Group
\end{minipage} & \begin{minipage}[b]{\linewidth}\raggedleft
Total n
\end{minipage} & \begin{minipage}[b]{\linewidth}\raggedleft
Sample share
\end{minipage} & \begin{minipage}[b]{\linewidth}\raggedleft
Reported AI-service use, item 0.6
\end{minipage} & \begin{minipage}[b]{\linewidth}\raggedleft
Weekly use, item 1.4
\end{minipage} \\
\midrule\noalign{}
\endhead
\bottomrule\noalign{}
\endlastfoot
Students & 1809 & 85.3\% & 1503/1809 = 83.08\% & 692/1571 = 44.05\% \\
Faculty & 250 & 11.8\% & 192/250 = 76.80\% & 46/213 = 21.60\% \\
Administrative staff & 62 & 2.9\% & 53/62 = 85.48\% & 9/57 = 15.79\% \\
Total & 2121 & 100.0\% & 1748/2121 = 82.41\% & Not applicable \\
\end{longtable}

\emph{Note.} Reported AI-service use was calculated using the full
analytical sample for each group. Weekly use was calculated only among
valid responses to item 1.4. When the full group is used as the
denominator, weekly use is 38.25\% among students, 18.40\% among
faculty, and 14.52\% among administrative staff.

\subsection{3.3. Instrument}\label{instrument}

The instrument comprised three role-adapted questionnaires for students,
faculty, and administrative staff. Each version contained 75 items. The
questionnaires had comparable block structures, but the wording was
adapted to learning, teaching, or administrative activities and was not
identical across versions.

The questionnaires covered the following main domains:

\begin{enumerate}
\def\labelenumi{\arabic{enumi}.}
\tightlist
\item
  Demographic context, AI-use experience, and competence.
\item
  Awareness of institutional policy and access to AI tools.
\item
  Current and expected frequency of AI use across several contexts.
\item
  Current and expected perceived usefulness.
\item
  Trust in AI and control over its contribution.
\item
  Policy clarity and institutional support.
\item
  Academic integrity and responsible-use norms.
\item
  Social norms, personal readiness, and contexts of permissible use.
\item
  Expected consequences of institutional scenarios.
\item
  Open-ended questions about practices, risks, and necessary changes.
\end{enumerate}

The questionnaires also included an optional contact item and two
attention-check items. Agreement items had five substantive response
categories, ranging from ``strongly disagree'' to ``strongly agree,'' as
well as the non-substantive options ``difficult to answer'' and ``prefer
not to answer.'' The latter options were treated as missing in scale
calculations. Frequency items were coded from 1 (``never'') to 5
(``several times per week/daily''). The present analysis used the full
cleaned datasets; attention-check results were not used as an exclusion
criterion.

\subsection{3.4. Variables and Scoring
Rules}\label{variables-and-scoring-rules}

The present analysis used role-specific items with the same numerical
suffixes and standard five-point coding.

\textbf{Current AI-use intensity} was calculated as the mean of
available valid responses to items 2.1-2.3, which measured generating
substantial portions of content, summarization, and idea generation.

\textbf{Perceived usefulness} was calculated from current items 3.1-3.3:
time savings, improved output quality, and support in understanding
material.

The \textbf{trust and control index} was calculated from items 4.1-4.3.
In the OLS trust model, the dependent variable was a two-item trust
score based on items 4.1 and 4.2, while item 4.3 was treated separately
as perceived control.

\textbf{Institutional policy clarity} was calculated from current items
5.1, 5.2, and 5.4. This index reflected the perceived clarity and
applicability of rules and was not a measure of awareness that a policy
exists, which was measured separately by item 1.1.

The \textbf{responsible AI-use norms index} was calculated from items
6A.1-6A.4. The \textbf{academic integrity concerns index} was calculated
from items 6A.3 and 6A.4. The second index was therefore an exact subset
of the first rather than an independent construct; their simultaneous
OLS coefficients may reflect overlap and suppression and should not be
substantively contrasted. \textbf{Perceived improvement in quality} in
the main group comparisons was current item 3.2 and was therefore
interpreted as a separate self-assessment, not an objective educational
outcome.

\textbf{AI-use experience} was coded from 0 (``do not use'') to 4
(``more than two years''), and \textbf{competence} from 1 (``complete
novice'') to 5 (``expert'').

For group descriptions and regression models, each index was first
calculated for each respondent as the mean of the available valid items.
A respondent could therefore have an index value without valid responses
to every component item. OLS then excluded respondents missing any
previously calculated model variable.

\subsection{3.5. Analytical Strategy}\label{analytical-strategy}

Frequencies, proportions, means, standard deviations, and 95\%
confidence intervals were calculated first. Given the substantial
imbalance in group sizes, differences in the study indices were tested
using Welch's one-way ANOVA. Pairwise comparisons used Welch t-tests
with a Bonferroni adjustment within each index. Classical eta squared
and Hedges' g were calculated as effect sizes.

Two pooled OLS models estimated associations with trust and current
AI-use intensity. Continuous dependent variables and predictors were
standardized. Both models included group indicators, with administrative
staff as the reference category, and were estimated on the same
subsample with complete data on all calculated model variables (\emph{N}
= 1254). Conventional OLS standard errors were used, and
predictor-by-group interactions were not estimated.

The reliability of multi-item indices was assessed using Cronbach's
alpha among respondents with complete data for the relevant item set. A
PCA-based variant of Harman's single-factor test was also conducted on
nine common items, followed by preliminary screening of cross-group
comparability using Tucker's congruence coefficient. Before the
one-factor loading estimates were calculated, missing item responses
were imputed with the within-group median for the relevant item. This
screening was not a formal test of measurement invariance.

Exploratory K-means clustering was conducted only among students. The
primary clustering specification used nine features: experience,
competence, three current usefulness items 3.1-3.3, expected future time
savings item 3.4, two trust items 4.1-4.2, and control item 4.3. The
complete-case subsample was \emph{n} = 961. Features were standardized,
after which a four-cluster solution (\emph{k} = 4) with a fixed random
seed of 42 and ten initializations was estimated. In this specification,
the two experience categories below one year were combined and
experience was coded from 0 to 3, while agreement responses were coded
as 1, 2, 3.5, 5, and 6. The cluster results were therefore treated
separately from the main indices and used only for exploratory analysis.

\subsection{3.6. AI-Assisted Manuscript
Preparation}\label{ai-assisted-manuscript-preparation}

Generative and AI-assisted tools supported translation, draft wording,
organization, and language and terminology review. The authors reviewed
and edited the content and take full responsibility for the manuscript.

\section{4. Results}\label{results}

\subsection{4.1. Overall Pattern of AI
Use}\label{overall-pattern-of-ai-use}

Respondents reported using AI in learning, teaching, and administrative
activities, but the prevalence of any reported AI-service use and the
rate of weekly use differed substantially. On item 0.6, 1748 of 2121
respondents (82.41\%) reported using at least one AI service. At the
group level, the prevalence of reported AI-service use was highest among
administrative staff (85.48\%), followed by students (83.08\%) and
faculty (76.80\%).

A different pattern emerged for weekly use. On item 1.4, the proportion
reporting AI use one to three times per week or more often among
respondents with valid data was 44.05\% for students, 21.60\% for
faculty, and 15.79\% for administrative staff. Thus, a high prevalence
of reported AI-service use does not imply equally high weekly use, and
the administrative group, which had the highest prevalence, had the
lowest weekly-use rate.

Role-adapted questionnaire items addressed different tasks: academic
writing and explanation of course material for students, preparation of
materials and feedback for faculty, and letters, documents, and routine
requests for administrative staff. Frequencies for these task contexts
and the open-ended responses were not analyzed here. These examples
therefore describe the content of the instrument rather than empirically
ranked practices.

\subsection{4.2. Reliability and Preliminary Measurement
Diagnostics}\label{reliability-and-preliminary-measurement-diagnostics}

Cronbach's alpha generally indicated acceptable internal consistency for
the main indices, although complete-case \emph{n} and reliability
differed across groups. Some estimates for the administrative group
should be interpreted with particular caution.

\textbf{Table 2. Cronbach's alpha and complete observations for the main
indices}

\begin{longtable}[]{@{}
  >{\raggedright\arraybackslash}p{(\linewidth - 6\tabcolsep) * \real{0.2000}}
  >{\raggedleft\arraybackslash}p{(\linewidth - 6\tabcolsep) * \real{0.2667}}
  >{\raggedleft\arraybackslash}p{(\linewidth - 6\tabcolsep) * \real{0.2667}}
  >{\raggedleft\arraybackslash}p{(\linewidth - 6\tabcolsep) * \real{0.2667}}@{}}
\toprule\noalign{}
\begin{minipage}[b]{\linewidth}\raggedright
Index
\end{minipage} & \begin{minipage}[b]{\linewidth}\raggedleft
Students \(\alpha\) (n)
\end{minipage} & \begin{minipage}[b]{\linewidth}\raggedleft
Faculty \(\alpha\) (n)
\end{minipage} & \begin{minipage}[b]{\linewidth}\raggedleft
Administrative staff \(\alpha\) (n)
\end{minipage} \\
\midrule\noalign{}
\endhead
\bottomrule\noalign{}
\endlastfoot
Current AI-use intensity & 0.866 (1437) & 0.859 (212) & 0.742 (54) \\
Perceived usefulness & 0.809 (1292) & 0.835 (158) & 0.920 (41) \\
Trust and control & 0.781 (1263) & 0.707 (141) & 0.807 (37) \\
Institutional policy clarity & 0.803 (871) & 0.846 (89) & 0.845 (28) \\
Responsible AI-use norms & 0.776 (874) & 0.734 (171) & 0.609 (42) \\
Academic integrity concerns & 0.778 (1017) & 0.789 (202) & 0.639 (46) \\
\end{longtable}

The PCA-based diagnostic showed that the first component explained
57.6\% of the total variance among students (\emph{n} = 755), 39.8\%
among faculty (\emph{n} = 68), 56.7\% among administrative staff
(\emph{n} = 25), and 56.2\% in the pooled complete-case sample (\emph{n}
= 848). This result indicates a need to consider the risk of common
method variance but is not a formal test of the entire questionnaire.

Tucker's congruence coefficients for one-factor loading patterns on the
same nine common items were 0.979 for students and faculty, 0.967 for
faculty and administrative staff, and 0.934 for students and
administrative staff. The lower value for students and administrative
staff further limits confidence in direct comparisons between these
groups. Overall, these values provide only preliminary evidence of
similar loading patterns, not confirmation of full measurement
invariance.

\subsection{4.3. Group Differences}\label{group-differences-1}

Students had the highest mean current AI-use intensity and perceived
usefulness. Faculty and administrative staff, by contrast, had higher
mean scores on responsible-use norms and academic integrity concerns.

\textbf{Table 3. Group comparisons of the study indices}

\begin{longtable}[]{@{}
  >{\raggedright\arraybackslash}p{(\linewidth - 12\tabcolsep) * \real{0.1111}}
  >{\raggedleft\arraybackslash}p{(\linewidth - 12\tabcolsep) * \real{0.1481}}
  >{\raggedleft\arraybackslash}p{(\linewidth - 12\tabcolsep) * \real{0.1481}}
  >{\raggedleft\arraybackslash}p{(\linewidth - 12\tabcolsep) * \real{0.1481}}
  >{\raggedleft\arraybackslash}p{(\linewidth - 12\tabcolsep) * \real{0.1481}}
  >{\raggedleft\arraybackslash}p{(\linewidth - 12\tabcolsep) * \real{0.1481}}
  >{\raggedleft\arraybackslash}p{(\linewidth - 12\tabcolsep) * \real{0.1481}}@{}}
\toprule\noalign{}
\begin{minipage}[b]{\linewidth}\raggedright
Index
\end{minipage} & \begin{minipage}[b]{\linewidth}\raggedleft
Students M (n)
\end{minipage} & \begin{minipage}[b]{\linewidth}\raggedleft
Faculty M (n)
\end{minipage} & \begin{minipage}[b]{\linewidth}\raggedleft
Administrative staff M (n)
\end{minipage} & \begin{minipage}[b]{\linewidth}\raggedleft
Welch F
\end{minipage} & \begin{minipage}[b]{\linewidth}\raggedleft
p
\end{minipage} & \begin{minipage}[b]{\linewidth}\raggedleft
eta²
\end{minipage} \\
\midrule\noalign{}
\endhead
\bottomrule\noalign{}
\endlastfoot
Current AI-use intensity & 2.73 (1650) & 1.84 (236) & 1.61 (58) & 128.92
& \textless0.001 & 0.083 \\
Perceived usefulness & 3.62 (1591) & 2.87 (209) & 3.08 (53) & 47.50 &
\textless0.001 & 0.057 \\
Trust and control & 3.61 (1537) & 3.11 (205) & 3.32 (55) & 22.65 &
\textless0.001 & 0.027 \\
Institutional policy clarity & 3.77 (1443) & 3.35 (172) & 3.45 (54) &
12.87 & \textless0.001 & 0.018 \\
Responsible AI-use norms & 3.47 (1452) & 4.17 (224) & 4.10 (59) & 79.61
& \textless0.001 & 0.065 \\
Academic integrity concerns & 3.40 (1317) & 4.38 (219) & 4.33 (56) &
122.66 & \textless0.001 & 0.086 \\
Perceived improvement in quality & 3.29 (1420) & 2.66 (182) & 2.98 (48)
& 20.89 & \textless0.001 & 0.026 \\
\end{longtable}

In pairwise comparisons, students had higher current AI-use intensity
than faculty (\emph{g} = 0.79, adjusted \emph{p} \textless{} 0.001) and
administrative staff (\emph{g} = 0.98, \emph{p} \textless{} 0.001). The
difference between faculty and administrative staff was not significant
after adjustment (\emph{p} = 0.149).

A similar pattern emerged for perceived usefulness: students had higher
values than faculty (\emph{g} = 0.75, \emph{p} \textless{} 0.001) and
administrative staff (\emph{g} = 0.54, \emph{p} = 0.006), while faculty
and administrative staff did not differ (\emph{p} = 0.742).

Faculty and administrative staff reported stronger academic integrity
concerns than students: \emph{g} = 0.81 for faculty versus students and
\emph{g} = 0.75 for administrative staff versus students, with \emph{p}
\textless{} 0.001 in both cases. Faculty and administrative staff did
not differ (\emph{p} = 1.000). Faculty and administrative staff also
showed higher mean scores on responsible-use norms than students.

Institutional policy clarity warrants particular attention. On the core
index of items 5.1, 5.2, and 5.4, students had a higher mean than
faculty (\emph{g} = 0.42, \emph{p} \textless{} 0.001). Neither the
adjusted difference between students and administrative staff (\emph{p}
= 0.147) nor that between faculty and administrative staff (\emph{p} =
1.000) was significant. Thus, students reported higher perceived policy
clarity than faculty, while neither group differed significantly from
administrative staff. This result should not be conflated with awareness
that a policy exists, which was measured separately by item 1.1.

\subsection{4.4. Regression Analysis}\label{regression-analysis}

The first pooled OLS model estimated associations with trust in AI and
explained 40.8\% of the variance in trust (\emph{N} = 1254, \emph{R}² =
0.408, adjusted \emph{R}² = 0.404). Perceived usefulness had the
strongest standardized positive association with trust (\(\beta\) =
0.402, \emph{p} \textless{} 0.001). Institutional policy clarity also
had a positive association (\(\beta\) = 0.223, \emph{p} \textless{}
0.001) but was not the strongest predictor.

\textbf{Table 4. Pooled OLS models of trust and current AI-use
intensity}

\begin{longtable}[]{@{}
  >{\raggedright\arraybackslash}p{(\linewidth - 6\tabcolsep) * \real{0.2143}}
  >{\raggedright\arraybackslash}p{(\linewidth - 6\tabcolsep) * \real{0.2143}}
  >{\raggedleft\arraybackslash}p{(\linewidth - 6\tabcolsep) * \real{0.2857}}
  >{\raggedleft\arraybackslash}p{(\linewidth - 6\tabcolsep) * \real{0.2857}}@{}}
\toprule\noalign{}
\begin{minipage}[b]{\linewidth}\raggedright
Outcome
\end{minipage} & \begin{minipage}[b]{\linewidth}\raggedright
Predictor
\end{minipage} & \begin{minipage}[b]{\linewidth}\raggedleft
Coefficient
\end{minipage} & \begin{minipage}[b]{\linewidth}\raggedleft
p
\end{minipage} \\
\midrule\noalign{}
\endhead
\bottomrule\noalign{}
\endlastfoot
Trust & Perceived usefulness & 0.402 & \textless0.001 \\
Trust & Institutional policy clarity & 0.223 & \textless0.001 \\
Trust & Responsible AI-use norms & 0.155 & \textless0.001 \\
Trust & Current AI-use intensity & 0.108 & \textless0.001 \\
Trust & Academic integrity concerns & -0.111 & 0.016 \\
Trust & AI-use experience & 0.026 & 0.354 \\
Trust & Competence & 0.037 & 0.203 \\
Trust & Faculty, ref. administrative staff & -0.288 & 0.040 \\
Trust & Students, ref. administrative staff & -0.128 & 0.322 \\
Current AI-use intensity & Perceived usefulness & 0.322 &
\textless0.001 \\
Current AI-use intensity & Competence & 0.268 & \textless0.001 \\
Current AI-use intensity & Trust & 0.106 & \textless0.001 \\
Current AI-use intensity & AI-use experience & 0.099 & \textless0.001 \\
Current AI-use intensity & Institutional policy clarity & -0.002 &
0.947 \\
Current AI-use intensity & Responsible AI-use norms & -0.032 & 0.492 \\
Current AI-use intensity & Academic integrity concerns & -0.023 &
0.612 \\
Current AI-use intensity & Faculty, ref. administrative staff & 0.211 &
0.129 \\
Current AI-use intensity & Students, ref. administrative staff & 0.545 &
\textless0.001 \\
\end{longtable}

\emph{Note.} Standardized coefficients \(\beta\) are reported for
continuous predictors. Coefficients for binary group indicators are
expressed in outcome standard deviations relative to the administrative
group. The \emph{p} values are based on conventional OLS standard
errors; intercepts are not shown. Because group interactions were not
estimated, the models assume equal slopes across groups.

The second model explained 42.1\% of the variance in current AI-use
intensity (\emph{N} = 1254, \emph{R}² = 0.421, adjusted \emph{R}² =
0.417). Perceived usefulness had the strongest positive association
(\(\beta\) = 0.322), followed by competence (\(\beta\) = 0.268), trust
(\(\beta\) = 0.106), and experience (\(\beta\) = 0.099). Policy clarity
was not associated with current AI-use intensity after adjustment for
the other variables (\(\beta\) = -0.002, \emph{p} = 0.947).

Coefficients in both models describe statistical associations among
self-reported measures. They are not directional causal effects and do
not constitute a latent-variable model. In particular, the coefficients
for responsible-use norms and academic integrity concerns cannot be
interpreted as independent effects because the two indices share items
6A.3 and 6A.4.

\subsection{4.5. Experience and Perceived
Usefulness}\label{experience-and-perceived-usefulness}

Among students, longer experience using AI was monotonically associated
with current perceived usefulness on the three-item index comprising
items 3.1-3.3. The mean increased from 2.78 among non-users (\emph{n} =
214) to 3.51 among those with less than six months of experience
(\emph{n} = 375), 3.81 among those with six to twelve months (\emph{n} =
328), 3.83 among those with one to two years (\emph{n} = 367), and 4.06
among those with more than two years (\emph{n} = 170). Spearman's rank
correlation was \emph{r} = 0.327, \emph{p} \textless{} 0.001, \emph{n} =
1454. Faculty showed a general increase followed by a plateau, whereas a
reliable trend could not be estimated for administrative staff because
only one respondent reported more than two years of experience.

This pattern is consistent with a positive association between
experience and perceived usefulness but does not establish that
accumulating experience causally increases perceived usefulness.
Self-selection is also possible: users who initially perceived AI as
more useful may have continued using it for longer.

\subsection{4.6. Exploratory Student
Profiles}\label{exploratory-student-profiles}

The primary student-only K-means specification identified four clusters
in the complete-case subsample of \emph{n} = 961. With the fixed random
seed and ten initializations described above, cluster sizes were 221,
111, 378, and 251.

Cluster 0 (\emph{n} = 221, 23.0\%) had moderate experience, usefulness,
and trust, with comparatively stronger control. Cluster 1 (\emph{n} =
111, 11.6\%) had the lowest experience, perceived usefulness, and trust.
Cluster 2 (\emph{n} = 378, 39.3\%) combined high perceived usefulness
and trust with moderate experience. Cluster 3 (\emph{n} = 251, 26.1\%)
had the highest experience, competence, perceived usefulness, and trust.

These clusters demonstrate heterogeneity within the student sample but
are not stable psychological types. The solution does not cover faculty
or administrative staff, is not a latent profile analysis, and is
sensitive to the feature set and parameters. A separate sensitivity
check used six aggregated features: experience, competence, usefulness
from items 3.1-3.3, trust from items 4.1-4.2, control item 4.3, and the
separate perceived improvement in quality item 3.2. Because item 3.2 was
already part of the usefulness index, it received additional weight in
this specification. In a different complete-case sample (\emph{n} =
1128), the largest silhouette value was obtained for \emph{k} = 2
(0.332), whereas the value for \emph{k} = 4 was 0.241. Because the two
specifications used different features and samples, this check cannot
show that the four-cluster solution is suboptimal. It does show,
however, that the optimality and stability of that solution have not
been established. The solution is therefore retained only as an
exploratory description of one specification.

The questionnaire did not measure a separate validated construct of
cognitive delegation. Cluster differences in trust in AI-generated logic
and structure can therefore be discussed only as indirect evidence
concerning reflective control, not as proof of willingness to delegate
cognitive effort.

\section{5. Discussion}\label{discussion}

The findings have several implications for theory, institutional policy,
and everyday educational practice.

First, AI can no longer be described solely as an ``external
innovation'' located outside the university. Most respondents reported
using at least one AI service, although this measure captured any
reported use rather than its frequency. Students reported weekly use
more often than faculty and administrative staff and perceived AI as
more useful. University policy should therefore address practices that
have already developed rather than rest primarily on assumptions about
future technology adoption.

Second, the study supports the idea that institutional policy clarity is
a governance resource for universities responding to AI. Perceived
clarity of rules was positively associated with trust in AI. However,
this association was not the strongest in the model estimated here:
perceived usefulness had the larger standardized coefficient. This
association should be interpreted alongside direct user experience,
competence, usefulness, and academic integrity concerns. The
cross-sectional data do not establish that policy clarity in itself
increases trust.

Third, faculty and administrative staff should not be viewed simply as
resistant to AI. Both groups reported AI use while also reporting
stronger academic integrity concerns and greater endorsement of
responsible-use norms than students. This more complex pattern indicates
that university policy should not rely on a blanket distinction between
permitted and prohibited uses. It requires a task-sensitive design, with
different rules for idea generation, text editing, material preparation,
assessment, fully AI-generated answers, and disclosure of AI use.

Fourth, the clustering results indicated heterogeneity within the
student sample in experience, competence, usefulness, trust, and
control. The exploratory clusters suggest that a single approach to
training and regulation may not serve inexperienced or skeptical users,
pragmatically positive users, and more experienced students equally
well. However, these clusters should not be treated as fixed
psychological categories or used to classify individual students.

Fifth, the findings also underscore the substantive role of
administrative staff. Administrative units are stakeholders in the
university AI environment in their own right: they connect policy with
everyday procedures, service regulations, internal guidance, and
workflows. The data do not show that administrative practices cause AI
adoption, but they are consistent with the need to consider
administrative processes alongside teaching and learning.

Conceptually, the study suggests a simple but productive interpretation:
perceived usefulness, academic integrity concerns, policy clarity, and
trust form an interconnected context for AI use. Benefits without clear
boundaries may coexist with normative uncertainty, whereas rules that
lack perceived practical value may have little effect in practice. These
propositions should be treated as interpretations of observed
associations and as a basis for further research, not as evidence of a
causal pathway to sustained institutional integration.

\section{6. Limitations}\label{limitations}

This study has several limitations. First, it was conducted within a
single university and does not claim cross-institutional
representativeness. Because the available records did not document the
sampling frame, recruitment channels, or response rate, selection and
nonresponse bias cannot be estimated.

Second, the study used a cross-sectional design; therefore, the observed
associations do not establish causal direction. They may reflect either
effects of experience on attitudes or user self-selection.

Third, the data were obtained from respondents' self-reports and may
therefore be affected by social desirability bias, especially on issues
related to academic integrity and inappropriate AI use. Attention checks
were not used to exclude responses from the present analysis. The
PCA-based diagnostic used a limited pool of common items and also
indicated that common method variance warrants consideration.

Fourth, group sizes differed substantially. The administrative sample
comprised 62 respondents; complete observations for the reliability of
individual indices ranged from 28 to 54 and fell to 25 for the nine-item
PCA diagnostic. This imbalance limits the precision of conclusions about
administrative staff and differences between the two staff groups.

Fifth, missing-data rules differed across analyses. Group means and OLS
indices were calculated from available valid items, after which OLS
excluded observations missing any calculated model variable. By
contrast, the primary K-means specification required complete responses
on all original features. Sample sizes therefore varied across analyses
and are reported with each result. The OLS subsample was complete on the
calculated index scores, not necessarily on every item that entered
those scores.

Sixth, Tucker congruence screening on nine common items with median
imputation of missing values was not a formal test of measurement
invariance. The value of 0.934 for students and administrative staff
requires particular caution. Group comparisons should be interpreted as
comparisons of role-adapted observed indices, not fully equivalent
latent constructs.

Seventh, the four-cluster K-means solution applied only to students,
used a separate coding scheme, and did not demonstrate optimality or
stability. It is not a latent classification.

Eighth, the responsible-use norms and academic integrity concerns
indices share items 6A.3 and 6A.4. Conventional pooled OLS standard
errors were not adjusted for possible heteroskedasticity, and group
interactions were not estimated. The individual coefficients of the
overlapping indices and differences in slopes across groups should
therefore not be interpreted substantively.

Ninth, the study did not use behavioral traces, learning logs, artifacts
from actual assignments, or objective measures of academic performance.
Perceived improvement in quality therefore reflects a subjective
assessment rather than a measured educational effect. The questionnaire
also did not include a separate cognitive-delegation scale.

Despite these limitations, the design is useful as a model for internal
university monitoring. It can identify group differences, track measures
of use, compare perceived usefulness with academic integrity concerns,
and formulate questions for subsequent longitudinal and behavioral
research.

\section{7. Conclusion}\label{conclusion}

This study examined AI use and attitudes among students, faculty, and
administrative staff at a single large university specializing in
teacher education. AI use was widespread but uneven across the
university environment. Students reported higher current AI-use
intensity and perceived usefulness, whereas faculty and administrative
staff reported stronger academic integrity concerns and greater
endorsement of responsible-use norms. Among students, longer experience
using AI was positively associated with perceived usefulness.

Institutional policy clarity was positively associated with trust in AI,
but perceived usefulness had the stronger standardized association in
the pooled OLS model. Students reported higher perceived policy clarity
than faculty, while neither group differed significantly from
administrative staff. These findings are consistent with
stakeholder-aware, task-sensitive institutional AI policies that address
academic integrity requirements and transparent disclosure of AI use.

These conclusions concern statistical associations and group differences
within one university sample. They do not demonstrate causal effects of
policy on trust, effects of AI on objective educational outcomes, or a
stable latent typology of all stakeholders in the university
environment.

\section{Appendix A. Structure of the Role-Adapted 75-Item
Questionnaires}\label{appendix-a.-structure-of-the-role-adapted-75-item-questionnaires}

Each respondent group received a role-adapted 75-item questionnaire with
a comparable block structure.

\begin{itemize}
\tightlist
\item
  Block 0. Demographics and AI-use experience: 7 items.
\item
  Block 1. AI policy awareness and access: 4 items.
\item
  Block 2. Current and expected frequency of AI use: 6 items.
\item
  Block 3. Current and expected perceived usefulness of AI: 6 items.
\item
  Block 4. Trust in AI and control, current and expected ratings: 6
  items.
\item
  Block 5. AI rules and institutional support: 8 items.
\item
  Block 6A. Academic integrity and responsible AI-use norms: 4 items.
\item
  Block 6B. Social norms in the respondent's environment: 4 items.
\item
  Block 6C. Personal readiness and future norms: 6 items.
\item
  Block 6X. Additional norms, including editing and disclosure of use: 3
  items.
\item
  Block 7. Curriculum links and AI-related skills: 4 items.
\item
  Block 8. Contexts of perceived permissibility and personal readiness:
  6 items.
\item
  Block 9. Expected consequences of institutional scenarios over 12-24
  months: 6 items.
\item
  Block 10. Open-ended questions: 2 items.
\item
  Block 11. Optional contact information for follow-up participation: 1
  item.
\item
  Block 12. Attention checks: 2 items.
\end{itemize}

\section{Declarations}\label{declarations}

\subsection{Availability of data and
materials}\label{availability-of-data-and-materials}

The anonymized datasets, codebook, role-adapted questionnaires, and
analysis scripts are available from the corresponding author on
reasonable request, subject to institutional requirements and safeguards
for small subgroups.

\subsection{Ethics approval and consent to
participate}\label{ethics-approval-and-consent-to-participate}

The university's senior administrative body (the rectorate) served as
the institutional approval body under the university's procedures for
research involving human participants and approved the study protocol
and ethics procedures. All participants were older than 16 years, and
informed consent was obtained before questionnaire completion. The
institution remains anonymized in this manuscript.

\subsection{Consent for publication}\label{consent-for-publication}

Not applicable.

\subsection{Competing interests}\label{competing-interests}

The authors declare that they have no competing interests.

\subsection{Funding}\label{funding}

This research received no specific grant from any funding agency in the
public, commercial, or not-for-profit sectors.

\subsection{Authors' contributions}\label{authors-contributions}

Yuriy S. Braun: Conceptualization, Methodology, Formal analysis,
Investigation, Writing -- original draft, Writing -- review \& editing,
Supervision. Salavat M. Khafizov: Investigation, Data curation,
Validation, Resources, Project administration, Writing -- review \&
editing. Both authors read and approved the final manuscript.

\subsection{Acknowledgements}\label{acknowledgements}

Not applicable.

\subsection{Use of generative AI and AI-assisted
technologies}\label{use-of-generative-ai-and-ai-assisted-technologies}

Generative and AI-assisted tools supported translation, draft wording,
organization, and language and terminology review. The authors reviewed
and edited the content and take full responsibility for the manuscript.

\section{References}\label{references}

Aleshkovskiy, I. A., Gasparishvili, A. T., Narbut, N. P., Krukhmaleva,
O. V., \& Savina, N. E. (2024). Russian students on the potential and
limitations of artificial intelligence in education. \emph{RUDN Journal
of Sociology, 24}(2), 335-353.
https://doi.org/10.22363/2313-2272-2024-24-2-335-353

Burneo-Arteaga, P., Lira, Y., Murzi, H., Balula, A., \& Costa, A. P.
(2025). Capability-based training framework for generative AI in higher
education. \emph{Frontiers in Education, 10}, Article 1594199.
https://doi.org/10.3389/feduc.2025.1594199

Buyakova, K. I., Dmitriev, Y. A., Ivanova, A. S., Feshchenko, A. V., \&
Yakovleva, K. I. (2024). Students' and teachers' attitudes towards the
use of tools with generative artificial intelligence at the university.
\emph{The Education and Science Journal, 26}(7), 160-193.
https://doi.org/10.17853/1994-5639-2024-7-160-193

Cotton, D. R. E., Cotton, P. A., \& Shipway, J. R. (2024). Chatting and
cheating: Ensuring academic integrity in the era of ChatGPT.
\emph{Innovations in Education and Teaching International, 61}(2),
228-239. https://doi.org/10.1080/14703297.2023.2190148

Hackl, V., Müller, A. E., \& Sailer, M. (2026). The AI literacy
heptagon: A structured approach to AI literacy in higher education.
\emph{Computers and Education: Artificial Intelligence, 10}, Article
100540. https://doi.org/10.1016/j.caeai.2026.100540

Hugerth, M. W., \& Hugerth, L. W. (2026). Seeking knowledge or
efficiency: Profiling students' AI-use through survey-based latent class
analysis. \emph{Computers and Education: Artificial Intelligence, 10},
Article 100531. https://doi.org/10.1016/j.caeai.2025.100531

Humble, N., \& Mozelius, P. (2026). Beyond the hype: How higher
education stakeholders view the benefits and concerns of generative AI
for teaching, research, and administration. \emph{Computers and
Education Open, 10}, Article 100381.
https://doi.org/10.1016/j.caeo.2026.100381

Kasneci, E., Sessler, K., Küchemann, S., Bannert, M., Dementieva, D.,
Fischer, F., Gasser, U., Groh, G., Günnemann, S., Hüllermeier, E.,
Krusche, S., Kutyniok, G., Michaeli, T., Nerdel, C., Pfeffer, J.,
Poquet, O., Sailer, M., Schmidt, A., Seidel, T., . . . Kasneci, G.
(2023). ChatGPT for good? On opportunities and challenges of large
language models for education. \emph{Learning and Individual
Differences, 103}, Article 102274.
https://doi.org/10.1016/j.lindif.2023.102274

Lai, C. Y., Cheung, K. Y., Chan, C. S., \& Law, K. K. (2024).
Integrating the adapted UTAUT model with moral obligation, trust and
perceived risk to predict ChatGPT adoption for assessment support: A
survey with students. \emph{Computers and Education: Artificial
Intelligence, 6}, Article 100246.
https://doi.org/10.1016/j.caeai.2024.100246

Mariñas, K. A., Saflor, C. S., Alvarado, P., Uminga, J. M., \& Verde, N.
A. (2025). Assessing the importance of variables from a revised
technology acceptance model for the use of ChatGPT by university
students. \emph{Computers and Education: Artificial Intelligence, 9},
Article 100435. https://doi.org/10.1016/j.caeai.2025.100435

Mo, F., Huang, J., Yang, Y., Özen, Z., Maeda, Y., \& Olenchak, F. R.
(2026). Undergraduate students' learning outcomes with ChatGPT: A
meta-analytic study. \emph{Computers and Education: Artificial
Intelligence, 10}, Article 100536.
https://doi.org/10.1016/j.caeai.2025.100536

Saihi, A., Ben-Daya, M., Hariga, M., \& As'ad, R. (2024). A structural
equation modeling analysis of generative AI chatbots adoption among
students and educators in higher education. \emph{Computers and
Education: Artificial Intelligence, 7}, Article 100274.
https://doi.org/10.1016/j.caeai.2024.100274

Stojanov, A., Liu, Q., \& Koh, J. H. L. (2024). University students'
self-reported reliance on ChatGPT for learning: A latent profile
analysis. \emph{Computers and Education: Artificial Intelligence, 6},
Article 100243. https://doi.org/10.1016/j.caeai.2024.100243

Sun, D., Ba, S., Cha, Y., Yu, J., Chiang, F.-K., Dai, H. M., \& Lim,
C.-P. (2026). Empowering university teachers in higher education: A
generative AI-responsive competency framework. \emph{Computers and
Education: Artificial Intelligence, 10}, Article 100542.
https://doi.org/10.1016/j.caeai.2026.100542

Sysoyev, P. V. (2024). Ethics and AI-plagiarism in an academic
environment: Students' understanding of compliance with author's ethics
and the problem of plagiarism in the process of interaction with
generative artificial intelligence. \emph{Higher Education in Russia,
33}(2), 31-53. https://doi.org/10.31992/0869-3617-2024-33-2-31-53

Tlili, A., Shehata, B., Adarkwah, M. A., Bozkurt, A., Hickey, D. T.,
Huang, R., \& Agyemang, B. (2023). What if the devil is my guardian
angel: ChatGPT as a case study of using chatbots in education.
\emph{Smart Learning Environments, 10}, Article 15.
https://doi.org/10.1186/s40561-023-00237-x

Torres-Díaz, J. C., Duart, J., Rivera, D., \& Flandoli, A. B. (2025).
Artificial intelligence and academic integrity: Exploring plagiarism in
Ecuadorian universities. \emph{International Journal for Educational
Integrity, 21}, Article 35. https://doi.org/10.1007/s40979-025-00209-3

Zawacki-Richter, O., Marín, V. I., Bond, M., \& Gouverneur, F. (2019).
Systematic review of research on artificial intelligence applications in
higher education: Where are the educators? \emph{International Journal
of Educational Technology in Higher Education, 16}, Article 39.
https://doi.org/10.1186/s41239-019-0171-0

\end{document}